\documentclass{aa}
\usepackage{graphics}

\def\aa{\AA}
\def\teff{$T_\mathrm{eff}$}
\def\k{$\mathrm{K}$}
\def\logg{$\log g$}

\def\gf{{\it gf}}
\def\kms{$\mathrm{km~s^{-1}}$}
\def\cms{$\mathrm{cm~s^{-1}}$}
\def\leps{$\mathrm{\log~\epsilon}$}
\def\alfa{$\alpha$}
\def\csi{$\xi$}
\def\eg{e.g.}
\def\ie{i.e.}

\begin{document}

\thesaurus{08.16.3; 08.01.1; 08.01.3; 08.09.3; 02.12.3}

\title{First UVES observations of beryllium in very metal-poor 
stars\thanks{Based on public data released from the UVES Commissioning 
at the VLT/Kueyen telescope, European Southern Observatory, Paranal, Chile}}

\author{
F. Primas\inst{1}
\and P. Molaro\inst{2}
\and P. Bonifacio\inst{2}
\and V. Hill\inst{1}
}

\offprints{F. Primas}

\institute{European Southern Observatory, Karl-Schwarzschild Str. 2, 
D-85748 Garching b. M\"{u}nchen
\and Osservatorio Astronomico di Trieste, Via G.B. Tiepolo 11, 
I-34100 Trieste
}

\date{Received ; Accepted}
\authorrunning{Primas et al.}
\titlerunning{First UVES Observations of Beryllium}
\maketitle

\begin{abstract}
We report on a attempt to detect beryllium in two dwarf stars of the 
Galactic halo with metallicities below one thousandth solar. The data 
were obtained during the Commissioning of the Ultraviolet and Visible 
Echelle Spectrograph (UVES) mounted on the ESO VLT Kueyen telescope, 
and show the potential of UVES for studies in the UV-optical domain. 
We claim a beryllium detection in LP 815--43 ([Fe/H]~$\sim~-$2.95, 
[Be/H]~=~$-$13.09) at the 99.7\% confidence level, while only an upper 
limit can be set for the second target CD$-$24\degr17504 ([Fe/H]~$\sim~
-$3.30, [Be/H]~$\leq~-$13.39). These results suggest that the trend of 
beryllium with metallicity keeps decreasing as lower metallicities are 
probed, with no evidence for flattening. In CD$-$24\degr17504 we also 
analyzed the \ion{Li}{i} line at 6708~\AA, and derived a lithium abundance 
close to the Spite plateau. 

\end{abstract}

\section{Introduction}
The rare light elements (Li, Be, and B) are probes of the early 
universe, Galactic evolution, and stellar structure. Beryllium has a 
special place in the general scheme of nucleosynthesis, being the lightest 
stable nuclide not synthesized in the Big Bang. Together with 
\element[][6]{Li} and \element[][10]{B}, it is considered a pure product of 
cosmic-ray (CR) spallation nucleosynthesis, being generated only by the 
bombardment of \element[][13]{C} and \element[][16]{O} by protons and 
\alfa-particles (Reeves et al. \cite{reeves70}; Meneguzzi et al. 
\cite{meneg71}). This unique origin has made it a particularly useful 
monitor of time-integrated factors of Galactic evolution such as the product 
of particle fluxes and abundance of targets, since its production during the 
Galactic epoch appears to be limited to the interstellar medium (ISM). 
Recent studies of Be in halo stars (\eg\ Molaro et al. \cite{pm97}, 
Boesgaard et al. \cite{bk99}) have suggested that the nucleosynthesis 
processes responsible for its formation may be more complex than previously 
supposed; the linearity observed in the trend [Be/H] vs. [Fe/H]\footnote{
[A/H]=log(A/H)$_*$/log(A/H)$_{\sun}$} cannot be easily reproduced by 
spallation reactions between \alfa-particles and protons hitting CNO in the 
ISM. Hence, the study of the evolution of Be in the Galaxy is an important 
constraint of Galactic cosmic-ray (GCR) theory. The above--mentioned 
linearity, in fact, seems to support the idea that Type~II supernovae (SN) 
accelerate freshly synthesized C and O and subsequently fragment into 
Be and B (Vangioni-Flam et al. \cite{vf98}). New data, especially at low 
metallicities (below [Fe/H]~=~$-$3.0), are essential to distinguish between 
different hypotheses, like, for instance, the mass interval of the SN 
progenitor. 

Although this linearity strongly suggests a Galactic origin for Be, some 
inhomogeneous Big Bang Nucleosynthesis models (IBBN) have shown to be able 
to produce beryllium abundances as high as log (Be/H)~=~$-$13.00 (Kajino 
\& Boyd \cite{kajino90}; cf Orito et al. \cite{orito97} for a 
more recent review), \ie\ potentially observable in very metal-deficient 
stars. Such Big Bang component may appear as a constant Be-plateau, 
independent of metallicity, similar to what is found for lithium (cf Spite 
\& Spite \cite{sspite1}), but beryllium has been analyzed in one star 
only (BD $-$13$^{\circ}$3442) at [Fe/H]$\sim-$3.0 (Boesgaard et al. 
\cite{bk99}). Thus, this hypothesis has not been fully discarded yet.  

Here, we report on our very recent attempt of measuring beryllium in two 
of the most metal-poor stars ever observed in the spectral region near the 
atmospheric cut-off, where the Be lines fall (around 3130~\AA). These two 
new measurements will be compared to the current observational picture and 
we will show that this type of observations and measurements are now well 
within reach of UVES at VLT. 

\begin{table*}
\caption{Log of the UVES observations.}
\begin{flushleft} 
\begin{tabular}{lcccccc} \hline 
Star			& V & Date & Setting & Slit Width & Exp. Time & 
S/N\\
			& mag & & & arcsec & sec & per pixel \\ \hline
LP 815-43		& 10.9 & Oct 16 & B346 & 0.8'' & 1x5400 & 75 \\ 
			&      & Oct 8  & B346 & 1.1'' & 2x2700 & 
65$^{\mathrm{a}}$ \\
			&      &        &      &       &  10800 & 
110$^{\mathrm{b}}$  \\
CD$-$24\degr17504	& 12.2 & Oct 9  & B346 & 1.0'' & 3x3000 &  
80$^{\mathrm{b}}$  \\ 
			&      & Oct 9  & R580 & 1.0'' & 1x3000 & 
200  \\
			&      & Oct 9  & R860 & 1.0'' & 1x3000 & 
170  \\ \hline
\end{tabular} 
\end{flushleft} \label{tab1}
\begin{list}{}{} 
\item[$^{\mathrm{a}}$] measured on each spectrum (always close to the 
\ion{Be}{ii} lines in the B346 setting)
\item[$^{\mathrm{b}}$] measured on the final summed spectrum (close to the 
\ion{Li}{i} line in both R580 and R860 settings) 
\end{list} 
\end{table*}

\section{Observations and Data Reduction}

Despite of the fact that the first attempt of measuring beryllium in halo 
stars dates back to more than fifteen years ago (cf Molaro \& Beckman 
\cite{mb84}), only in the past few years have technological improvements 
allowed us to obtain high resolution and high signal-to-noise (S/N) 
ratios to analyze the Be resonance doublet features in the near-UV. 

The combination of atmospheric extinction, line crowding, and weakness 
of the \ion{Be}{ii} lines as lower metallicity stars are probed have made 
spectroscopic observations of beryllium very challenging when efficient 
near-UV detectors coupled to high-resolution spectrographs were not 
available. The first remarkable improvement in this respect (compared to 
the 4m class telescopes and instruments used at the beginning of the first 
extensive analyses of beryllium in metal-poor stars) is represented by the 
10m Keck~I telescope equipped with HIRES (Vogt et al. \cite{vogt94}), that 
permitted the determination of the Be abundances to a much higher 
accuracy than ever done before. BD $-$13$^{\circ}$3442, the most metal-poor 
star ([Fe/H]~=~$-$3.0) for which a Be detection has been claimed (Boesgaard et 
al. \cite{bk99}) is a V=10.3 star, which required 11 hours of integration 
time with HIRES at Keck~I in order to reach a S/N$\sim$130 at 3130~\AA.  
 
The  high-resolution Ultraviolet and Visible Echelle Spectrograph, designed 
and built at the European Southern Observatory, and that has now been 
mounted at one of the two Nasmyth foci of the second VLT 8m telescope 
(Kueyen), represents the most recent instrumental achievement related to this 
specific field of research. The first Commissioning period of UVES took 
place between September 27 and October 17, 1999, and all the scientific data 
have now become publicly available to the ESO community. The near-UV 
spectral region was observed in two very metal-deficient objects, LP 815--43 
and CD$-$24\degr17504 ([Fe/H]~=~$-$3.05 and $-$3.55 respectively, as found in 
the literature). The log of the observations and the spectra analyzed in this 
study are summarized in Table~1. 

Standard tasks of the Echelle reduction package in IRAF were applied to both 
sets of spectra. The first step included order definition, subtraction of 
bias and background between the orders, and flat-fielding. Because of the 
extreme weakness of the Be lines, we further checked the spectral region 
of the \ion{Be}{ii} resonance doublet in the non-flatfielded spectra, which 
confirmed the same shape of the lines and eliminated the possibility that 
spurious effects might had been introduced during the flat-fielding process. 
Subsequently, the orders were extracted and wavelength calibrated. This 
calibration achieved an {\it rms} deviation for a 2-dimensional 3rd order 
fit between pixel and wavelength space of 0.0016~\aa. The orders were then 
normalized to a continuum of 1, via a fitting procedure ({\rm continuum} 
task) with a spline of the 5th order. 

The resolution in the near-UV, as measured from the Full Width Half Maximum 
(FWHM) of the Th lines, is 48\,000 and 40\,000 (for the 2 sets of LP 815--43 
spectra, which were taken with different slit widths), and 40\,000 for the 
CD$-$24\degr17504 star. 

\begin{figure}[t] 
\vspace{-4cm}
\resizebox{9.2cm}{11cm}{\includegraphics{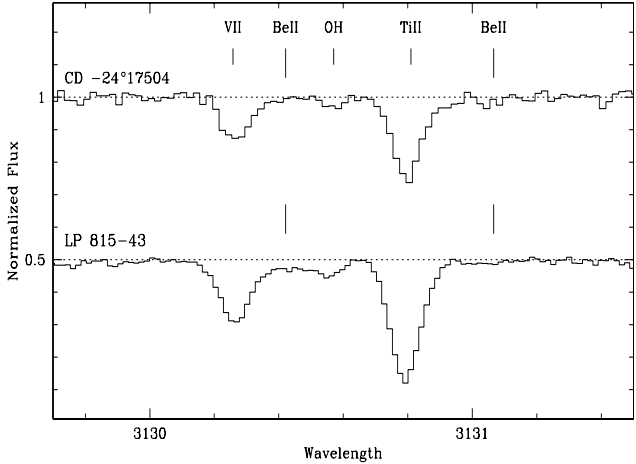}}
\caption{Reduced and normalized observed spectra of the two 
program stars (LP 815--43 has been shifted by $-$0.5 
in the y-coordinate)}
\label{fig1} 
\end{figure}

\begin{figure}[t]
\vspace{-4cm}
\resizebox{9.2cm}{11cm}{\includegraphics{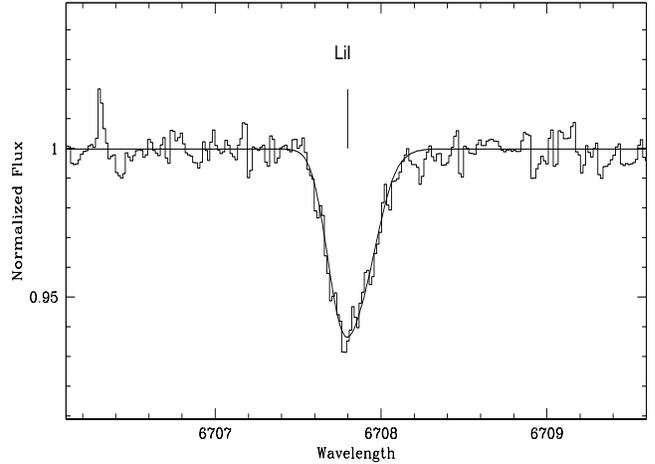}}
\caption{The \ion{Li}{i} $\lambda$6708 region in CD$-$24\degr17504 
(histogram). The thin line represents a synthetic spectrum computed with an 
ATLAS9 no-overshooting $\alpha$ enhanced model (\teff~=~6373~K, \logg~=~4.5, 
[M/H]~=~$-$3.5).}
\label{fig2} 
\end{figure}

All the spectra were then registered for radial velocity shifts and combined 
via a weighted sum (weighted by the inverse of the S/N squared of each 
spectrum). In the case of LP 815--43, because of the above--mentioned 
difference in spectral resolution (see Table~1), the highest resolution 
spectrum was degraded to the resolution of the second set of data by 
convolving it with a Gaussian function, before summing the spectra. 
Fig.~\ref{fig1} shows the beryllium spectral region in both stars and 
Fig.~\ref{fig2} shows the Li 6708~\AA~feature in CD$-$24\degr17504, at the 
end of the reduction procedure.

\section{Stellar Parameters}

Neither star has been widely studied in the literature. Ryan et al. 
(\cite{ryan91}, \cite{ryan99}) performed an extensive chemical abundance 
analysis and derived the following stellar parameters: \teff~=~$6340 \pm 
30$~\k~(an average value of several colour-temperature calibrations), 
\logg~=~4.0~dex (based on the ionization balance of \ion{Fe}{i} and 
\ion{Fe}{ii} lines), [Fe/H]~=~$-$3.05~dex (assuming \leps(Fe)$_{\odot}$
=~7.50), \csi~=~2.0~\kms~for LP 815--43, and \teff~= $6070 \pm 20$~\k, 
\logg~=~4.0~dex (which in this case was {\it assumed}), [Fe/H]~=~$-$3.55~dex, 
\csi~=~1.5~\kms~for CD$-$24\degr17504. From the fitting of the H\alfa~wings, 
Spite et al. (\cite{sspite3}) determined \teff~=~6300~\k~for both objects. 

In order to finalize our choice of stellar parameters, we decided to take 
advantage of the colour information available from Ryan et al. 
(\cite{ryan99}; cf Table~2, this work). We used both Carney (\cite{car83}) 
and King (\cite{king93}) \teff~{\it vs.} (b-y) calibrations, and derived 
respectively \teff~=~6501.52~\k~and 6527.34~\k~in the case of LP 815--43 and 
\teff~=~6187.46~\k~and 6287.31~\k~for CD$-$24\degr17504. We note that both 
stars have recent (unpublished) JHK photometry, to which a direct application 
of the InfraRed Flux Method (IRFM) provides \teff~=~6557~\k~for LP 815--43 
and \teff~=~6373~\k~for CD$-$24\degr17504 (Alonso, {\it private 
communication}). 

\begin{table*} 
\caption{Colour information and adopted stellar parameters.}
\begin{flushleft} 
\begin{tabular}{lccccccc} \hline 
Star & (b-y) & c$_1$ & E(b-y) & \teff & \logg & [Fe/H] & \csi \\ 
     &       &       &        &  K    & \cms  & dex    & \kms \\ \hline
LP 815--43 & 0.304 & 0.382 & 0.033 & 6500 & 4.25 & $-$2.95 & 1.75 \\
CD$-$24\degr17504 & 0.322 & 0.283 & 0.015 & 6300 & 4.50 & $-$3.30 & 1.00 \\ 
\hline
\end{tabular}
\end{flushleft}
\end{table*}

As far as the gravity is concerned, we initially assumed \logg~=~4.0, as 
suggested by Ryan and collaborators. From a quick inspection of the position 
of our targets in the evolutionary diagram c$_1$ {\it vs} (b-y) (which gives 
information on the evolutionary status of the object) compared to the 
Schuster \& Nissen (\cite{sn89}) loci used as reference, we found that 
gravities lower than 4.0 could be excluded (the stars fall very close to the 
turn-off, if not still on the main sequence). A cross-check with the 
isochrones of Bergbusch \& Vandenberg (\cite{bvdb92}) and Vandenberg \& 
Bell (\cite{vbb85}) provided consistent information: gravities slightly 
higher than 4.0 (4.35 and 4.45 respectively) were derived when an age of 
14~Gyr (although no difference was detected between 12, 14, and 16~Gyr) 
and the most metal-poor isochrone (which corresponds to [Fe/H]~=~$-$2.26) are 
assumed. These gravity values correspond to \teff~=~6560~K (LP 815--43) and 
\teff~=~6300~K (CD$-$24\degr17504). Because beryllium is strongly dependent 
on the choice of gravity, we decided to further check \logg~via the ionization 
balance. For this purpose, several lines of both titanium and iron in two 
different ionization stages (neutral and ionized) were selected between 3100 
and 3800~\AA. Their oscillator strengths were taken from the latest works of 
Martin et al. (\cite{martin88}) and Fuhr et al. (\cite{fuhr88}), and in the 
case of neutral iron were further cross--checked with the compilation of 
Nave et al. (\cite{nave94}). The accuracy given in these compilations (from 
A to D, \ie\ from 10 to 50\%) drove the final selection of the subsample of 
lines that were then used to check the ionization balances (no ``D'' line 
was used, and only few ``C''). The first run of WIDTH9 (Kurucz 
\cite{kur93}) was performed assuming \teff~=~6500~\k, \logg~=~4.0, 
[Fe/H]~=~$-$3.0 for LP 815--43, and \teff~=~6250~\k, \logg~=~4.0, 
[Fe/H]~=~$-$3.5 for CD$-$24\degr17504 respectively. First, by requiring no 
dependence of the abundance on the equivalent width, the microturbulence was 
constrained to 1$.75 \pm 0.2$~\kms~and $1.0 \pm 0.2$~\kms~for LP 815--43 and 
CD$-$24\degr17504 respectively. Then, the same code was run for different 
values of gravity ($\pm$~0.25, $\pm$~0.5) and temperature ($\pm$~250~K). The 
ionization balance was checked by using (10\ion{Fe}{i},~8\ion{Fe}{ii}) and 
(5\ion{Ti}{i},~13\ion{Ti}{ii}) lines for LP 815--43, and (6\ion{Fe}{i},~7
\ion{Fe}{ii}) plus (2\ion{Ti}{i},~13\ion{Ti}{ii}) for CD$-$24\degr17504 (cf 
Table~3, where LP and CD stands for LP 815--43 and CD$-$24\degr17504 
respectively); slightly higher values (4.25 and 4.5) were found confirming 
what we had derived from the isochrones. These are in good agreement, within 
the errors, with the values determined by Thevenin \& Idiart (\cite{thev99}), 
who studied non Local Thermodynamic Equilibrium (NLTE) corrections for iron 
abundances, and found \logg~=~4.39 for both stars. 

Considering all the estimates of \teff~and \logg~thus obtained and their 
quite good agreement, we adopted \teff~=~6500~K, \logg~=~4.25, and 
\csi~=~1.75~\kms~in the case of LP 815--43, and \teff~=~6300~K, \logg~=~4.5, 
and \csi~=~1.0~\kms~for CD$-$24\degr17504 as our final stellar parameters. 

Metallicities, on the contrary, were taken from the work of Ryan et al. 
(\cite{ryan99}), but corrected for the different temperature and gravity 
we adopted. The values thus found, [Fe/H]~=~$-$2.90 and $-$3.32 for LP 815--43 
and CD$-$24\degr17504 respectively, are in very good agreement with the 
metallicity inferred from our spectrum synthesis analysis (see next 
section).

By evaluating the uncertainties in the different methods followed 
to determine the stellar parameters, we find that $\pm$100~\k~in \teff, 
$\pm$~0.25 in \logg, $\pm$0.15~dex in metallicity, and $\pm$~0.2\kms~in the 
microturbulent velocity are representative of the uncertainty associated to 
each single parameter.  

\section{Spectrum Synthesis Analysis}

Our spectrum synthesis calculations were performed with the Kurucz' grid of 
model atmospheres and synthesis codes ATLAS and SYNTHE (Kurucz 1993, 
officially released on CD-ROMs). The model atmospheres were computed 
according to the above--mentioned adopted stellar parameters. Solar abundances 
were taken from the compilation of Anders \& Grevesse (\cite{ag89}). We used 
the line list previously tested by Primas et al. (\cite{fp97}) and the 
abundance ratios determined by Ryan et al. (\cite{ryan91}) in order to 
constrain the allowed variations of the elemental abundances that play a 
role in this part of the spectrum (e.g. titanium, vanadium and chromium). 

\begin{table}[t] 
\caption{Atomic Data.}
\begin{flushleft} 
\begin{tabular}{crcc} \hline 
Lambda  & log{\it gf} & E.W.$_{LP}$ & E.W.$_{CD}$ \\ 
\AA	&	   & m\AA & m\AA \\ \hline
\ion{Ti}{i}	&	   &      &      \\
3186.45 & $-$0.069 & 4.4  & ...  \\
3191.99 & 0.068	   & 11.4 & ...  \\
3199.92 & 0.201	   & 11.8 & ...  \\
3635.46 & 0.048	   & 5.2  & 4.6  \\
3653.49 & 0.220    & 9.9  & 5.9  \\
\ion{Ti}{ii}&	   &      &      \\
3148.05 & $-$1.184 & 40.6 & 19.5 \\
3152.26 & $-$1.075 & 42.7 & 29.1 \\ 
3154.21 & $-$1.183 & 43.7 & 25.6 \\ 
3155.68 & $-$1.053 & 40.5 & 19.7 \\
3161.22 & $-$0.752 & 55.4 & 35.4 \\ 
3161.78 & $-$0.559 & 62.2 & 43.9 \\
3162.57 & $-$0.455 & 65.9 & 52.1 \\
3217.06 & $-$0.581 & 71.4 & 55.3 \\  
3234.52 & 0.336    & 109.3& 85.8 \\
3236.58 & 0.145    & 90.8 & 73.0 \\
3239.05 & $-$0.031 & 84.5 & 55.8 \\
3241.99 & $-$0.136 & 84.7 & 66.0 \\
3251.92 & $-$0.669 & 68.6 & 47.8 \\
\ion{Fe}{i}	&	   &      &      \\
3175.45 & $-$0.620 & 11.7 & ...  \\
3193.23 & $-$2.220 & 33.9 & ...  \\
3199.53 & $-$0.510 & 12.0 & 10.5 \\
3442.36 & $-$1.393 & 3.6  & ...  \\
3469.83 & $-$1.633 & 2.6: & ...  \\
3495.29 & $-$0.920 &  8.9 & ...  \\
3570.09 & 0.153    & 92.7 & 66.6 \\
3608.86 & $-$0.100 & 70.0 & 60.0 \\
3631.46 & $-$0.036 & 69.8 & 61.8 \\
3758.23 & $-$0.027 & 83.9 & 70.0 \\
3763.79 & $-$0.238 & 71.8 & 61.0 \\
\ion{Fe}{ii}	&	   &      &      \\
3183.11 & $-$2.100 & 33.4 & 17.1 \\
3186.74 & $-$1.670 & 46.8 & 30.8 \\
3192.91 & $-$1.950 & 49.2 & 20.9 \\
3196.07 & $-$1.730 & 46.6 & ...  \\
3210.44 & $-$1.690 & 48.5 & 28.6 \\
3213.31 & $-$1.270 & 62.5 & 44.7 \\
3227.74 & $-$1.060 & 79.1 & 59.9 \\
3277.35 & $-$2.300 & 49.8 & 32.8 \\ \hline
\end{tabular}
\end{flushleft}
\end{table}

\begin{figure}[t]
\vspace{-0.43cm}
\resizebox{9.2cm}{13cm}{\includegraphics{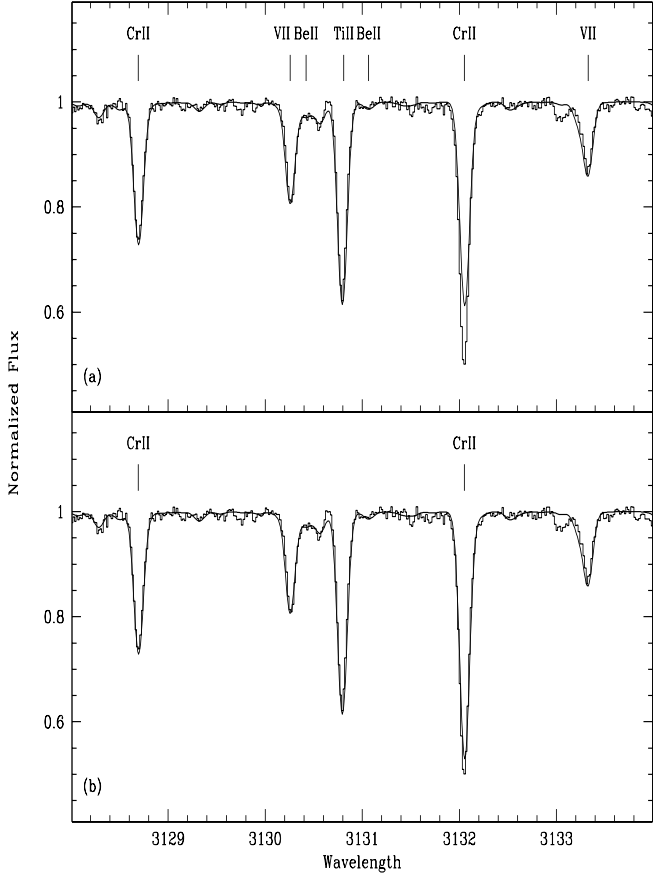}}
\caption{The region around the \ion{Be}{ii} resonance doublet in LP 815--43 
is shown (histogram). Surrounding lines of major interest are marked for easy 
identification. The thin line represents our best-fit synthesis. Panel (b) 
shows a better agreement in the fit of the two Cr lines (see text for more 
explanations)}
\label{fig3} 
\end{figure}

\begin{figure}[t]
\vspace{-3.6cm}
\resizebox{9.2cm}{10cm}{\includegraphics{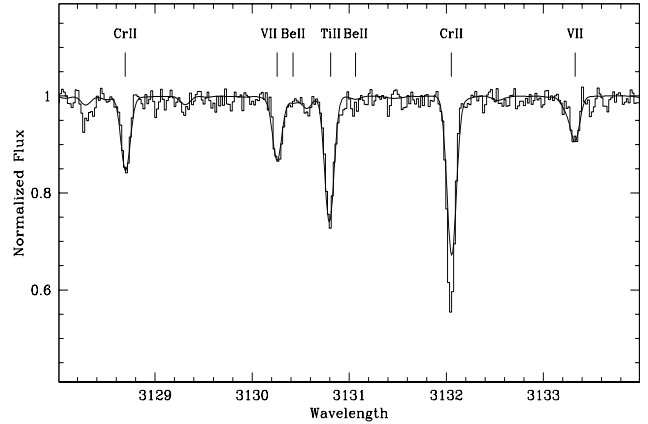}}
\caption{Same as Fig.~3, but for CD$-$24\degr17504}
\label{fig4} 
\end{figure}

Several syntheses were run until the best match between computed and 
observed spectra was achieved. Our best-fit syntheses were obtained with 
Kurucz' \alfa-enhanced (\ie~[\alfa/Fe]~=~+0.4~dex) model atmospheres, that 
also include the approximate overshooting. This choice was driven by the 
need of comparing in a consistent way our results with data available from 
the literature (\eg~Boesgaard et al. \cite{bk99}). The difference in the 
beryllium abundances derived by using models computed without the approximate 
overshooting was found to be negligible (on the order of ~0.05~dex). Our 
best-fit syntheses are shown in Fig.~\ref{fig3} and Fig.~\ref{fig4}, with 
all the relevant lines identified. They were obtained for \teff~=~6500~K, 
\logg~=~4.25, \csi~=~1.75~\kms, and [Fe/H]~=~$-$2.95 for LP 815--43, and 
\teff~=~6300~K, \logg~=~4.5, \csi~=~1.0~\kms, and [Fe/H]~=~$-$3.3 for 
CD$-$24\degr17504. Beryllium abundances equal to [Be/H]~=~$-$13.09 and 
$-$13.39 were adopted. Once the effect of using higher effective temperatures 
and gravities (compared to Ryan et al. \cite{ryan91}, \cite{ryan99}) on 
metallicity is considered, the metallicities we determine are in very good 
agreement with those of Ryan and collaborators. 

\begin{figure}
\vspace{-0.43cm} 
\resizebox{9.2cm}{13cm}{\includegraphics{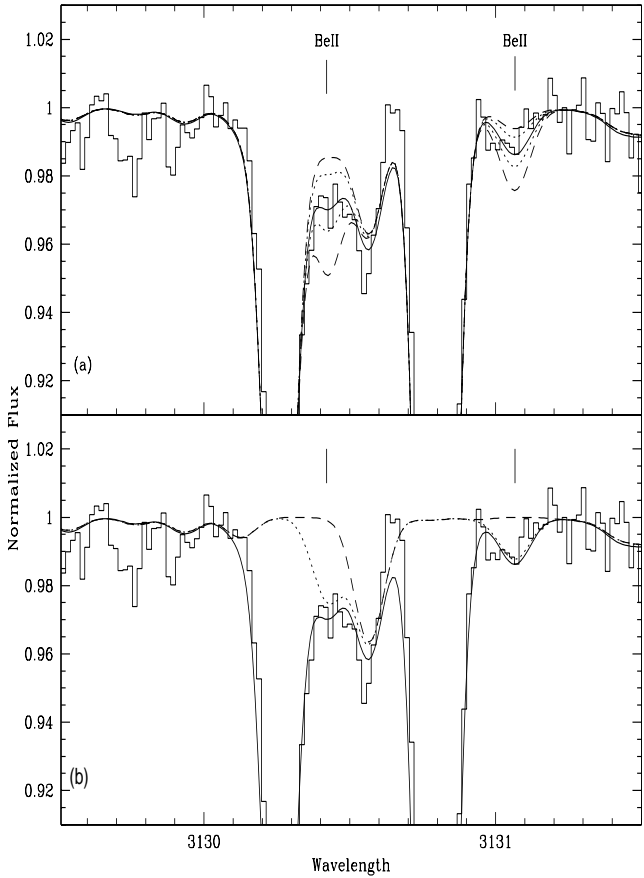}}
\caption{An enlarged view of the \ion{Be}{ii} resonance doublet in LP 815--43 
(histogram), together with our best-fit synthesis (thin line). Overplotted 
are: {\it (a)} four syntheses computed with the beryllium abundance increased 
and reduced respectively by 0.15~dex (dotted lines) and 0.30~dex (dashed 
lines); {\it (b)} one synthesis computed with our best-fit Be abundance and 
without the most important blending lines (dotted line) and one synthesis 
computed with no Be and no blending lines (dashed line)}
\label{fig5} 
\end{figure}

The main source of uncertainty affecting Be measurements comes from 
the accuracy with which the surface gravity is known for the stars under 
investigation and from the uncertainty related to the placement of the 
continuum. The dependence of Be on gravity is of the order of 
$\pm$~0.11--0.12~dex for a change in \logg~of $\pm$~0.25. An uncertainty of 
the order of 1--2\% in the determination of the continuum translates into 
another $\pm$~0.10dex. Altogether, these sum up to $\pm$~0.15~dex. However, 
inspection of Fig.~\ref{fig5}a tells us that $\pm$~0.15~dex may underestimate 
the total uncertainty to be associated with our measurements. Therefore, we 
decided to adopt $\pm$~0.20~dex as our representative error bar. 

Non-LTE effects on Be abundances have been discussed by Chmielewski 
et al. (\cite{chm75}), Kiselman \& Carlsson (\cite{kis94}) and Garc{\`\i}a 
L{\`o}pez et al. (\cite{garcia95}). Net NLTE corrections for halo dwarfs and 
other low mass stars are small, thus they were not included in the final 
estimate of the uncertainties.  

The only disagreement which emerged from the spectrum synthesis, common to 
both stars, concerns the abundance of chromium: the two lines present in 
this small spectral region cannot be fitted simultaneously (see 
Fig.~\ref{fig3}a and Fig.~\ref{fig4}). The solution of such disparity is 
beyond the scope of this contribution, but probably suggests some uncertainty 
in the \gf-values of these two lines. In the most recent works on transition 
probabilities, the log(\gf) value of the line at 3128.7~\aa ~is given an 
accuracy of "D" (50\%), and the redder line does not even appear (cf NIST, 
National Institute of Standards \& Technology). As a test, we changed 
(lowered) the \gf-value of the bluer line by different amounts (up to the 
allowed $\pm$50\%) and increased the Cr abundance up to 0.3~dex trying to 
find an optimal match. We were only partially successful in this exercise, 
having found a very good fit but for LP 815--43 only (see Fig.~\ref{fig3}b). 
Because of the lack of knowledge of the accuracy of the \gf-value of the 
redder \ion{Cr}{ii} line (3132.06~\aa), no conclusion can be drawn at this \
point. 

\section{The Beryllium Abundance and Its Implications}

When the lines of interest are very weak, as in our case, the abundances 
inferred from the spectrum syntheses can be usually interpreted as just 
consistent with the observed data. In order to check the validity of our 
Be detection, we applied Cayrel's formula (Cayrel \cite{cay88}), which 
estimates the minimum equivalent width detectable of a line, given the 
resolution, pixel size and signal-to-noise ratio of the observed data. In the 
case of our best spectrum (LP 815--43) and adopting a FWHM~=~0.1078~m\AA~and 
S/N~=~110 (at 3130~\AA), EW$_{min}$(1$\sigma$)~=~0.6~m\AA~is found. 
Unfortunately, even for our best spectrum, it is very difficult to measure 
the equivalent width of the feature present at 3131.066~\AA, that we think to 
be beryllium. Our attempt gives EW~=~1.7~m\AA, thus implying that we may be 
very close to a 3$\sigma$ detection. However, this procedure is not 
satisfactory.   

Following the referee's suggestion, we also tried to subtract the Gaussian 
fit of the main features blending with the two \ion{Be}{ii} lines, and then 
to fit the presumed \ion{Be}{ii} lines. This should provide a further check 
on the wavelength of the \ion{Be}{ii} lines. The test was indeed successful, 
and proved that the remaining lines fall exactly at 3130.421~\AA~and at 
3131.066~\AA, i.e. at the expected positions. However, we were not completely 
satisfied with this procedure either because the subtraction of the 
contribution due to the blending lines is a very delicate task (especially 
in very metal-poor stars where the \ion{Be}{ii} doublet is very weak and the 
blending features are still quite strong). 

A more compelling evidence that we have a detection in LP 815--43 may come 
from inspecting Fig.~\ref{fig5}b. After having checked that the absorptions we 
attribute to beryllium in LP 815--43 fall at the expected wavelengths, we 
tried to obtain a better estimate of the importance of the blending features. 
For this purpose, we ran two more syntheses: both of them were computed 
taking out all the features blending with the two \ion{Be}{ii} lines, but one 
had our previously determined best-fit Be abundance (dotted line in the 
figure) and the other had no beryllium (dashed line), i.e. the beryllium 
abundance was lowered by a factor of 20. In our opinion, this test shows 
that the blending features do not affect our determination of Be in this 
star, and that the features we initially attributed to beryllium are indeed 
the beryllium doublet. 

The observed spectrum of CD$-$24\degr17504 has a lower S/N$\sim$80, that 
makes the measurement of the equivalent width even more difficult and highly 
uncertain. The Be abundance derived for this star from our spectrum synthesis 
([Be/H]~=~$-$13.39) should be strictly considered an upper limit only. An 
observing strategy optimized for beryllium would likely have provided 
a detection. 

\begin{figure}[t]
\vspace{-4.77cm} 
\resizebox{9.2cm}{13cm}{\includegraphics{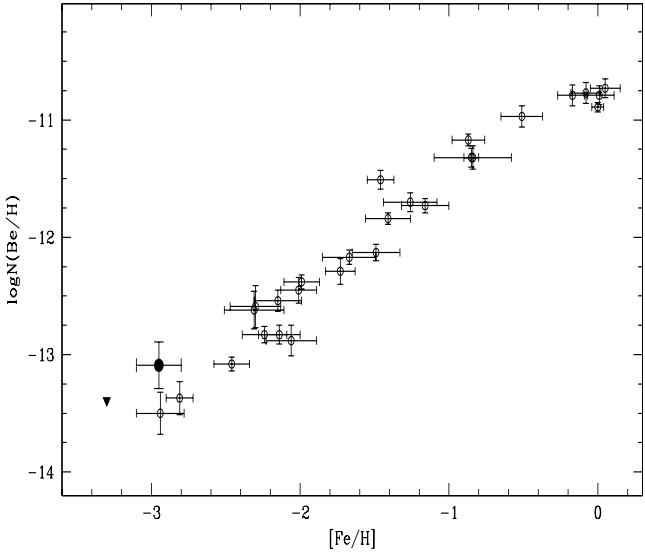}}
\caption{logN(Be/H) vs. [Fe/H] for the highest quality data points 
currently available. Open circles represent the work from Boesgaard et al. 
(1999) whereas the filled symbols represent this work (the circle is LP 
815--43 and the upside-down triangle is the upper limit for CD$-$24\degr17504)}
\vspace{-0.4cm}
\label{fig6} 
\end{figure}

Fig.~\ref{fig6} shows our two new Be results (filled circles) compared to the 
sample analyzed by Boesgaard et al. (\cite{bk99}, open circles -- Be 
abundances determined on the King \teff~scale), which represents the 
highest quality Be spectra available at the moment (the abundances were 
derived from high resolution, high S/N Keck~I HIRES spectra). Our two new 
data points (one detection, one upper limit) suggest that Be keeps decreasing 
as lower metallicities are probed, which is in support of a Galactic 
production of beryllium and argues against a primordial (Big Bang) component 
(although the latter cannot be excluded yet). 

The correlation of Be abundances with metallicity in the early Galaxy 
(Boesgaard et al. \cite{bk99}) and the finding of a B/Be ratio equal to 
that predicted by spallation (between 10 and 20, e.g. Duncan et al. 
\cite{dkd97}; Garc{\`\i}a L{\`o}pez et al. \cite{gl98}) have been usually 
considered a clear evidence for a Galactic (as opposed to primordial) 
production mechanism. Unfortunately, no B measurement is available for 
either of our two stars. This, together with the lack of Be determinations 
in stars below a metallicity of 1/1000 solar has so far prevented us from 
testing the efficiency of cosmic-ray spallation versus any possible 
primordial Be synthesis, as predicted by some Big Bang nucleosynthesis models 
that take into account inhomogeneities during the first few minutes after 
the Big Bang. The detection of a plateau in the relationship between 
[Be/H] and [O/H] (or [Be/H] vs [Fe/H], although less stringent) has usually 
been considered a possible evidence for such primordial abundance of Be, 
although the level of such plateau has remained quite uncertain (recent 
calculations, e.g. Orito et al. (\cite{orito97}), predict $\log$ N(Be/H) + 
12~=~$-$3.0, thus below the abundances we have been able to detect so far).  

However, should a plateau be detected, the interpretation may not be 
straightforward. Some of the theoretical scenarios recently proposed 
predict a Be plateau below [Fe/H]~=~$-$3.0 that is not correlated with a 
primordial production of beryllium. Yoshii et al. (\cite{yoshii95}) suggest 
that the finding of a Be plateau at low metallicities may derive from 
accretion phenomena of interstellar matter. These authors analyze how the 
accretion of metal-enriched interstellar gas onto metal-poor halo stars, 
while crossing the Galactic plane, may have affected the observed surface 
abundances of the light elements, beryllium and boron. 

According to their 
scenario, a key parameter that may distinguish between a primordial 
production of beryllium and the accretion scenario is the B/Be ratio 
(independent of the accretion rate, but strongly dependent on the baryon 
density in low- and high-density regions). Determinations of both Be and 
B in the same stars and in the region of the plateau become then a high 
priority in order to fully test the accretion scenario. But a similar 
plateau may also be related to the progenitor mass of the exploding 
supernova. Vangioni-Flam et al. (\cite{vf98}) proposed two possible 
different scenarios responsible for Be production. If shock acceleration 
in the gaseous phase of superbubbles produced by collective SN~II explosions 
is the main mechanism, then only the most massive stars (with initial mass 
M $\geq$ 60M$_{\odot}$) can play a role because of their much shorter 
lifetimes. Whereas, a much larger mass range (M $\geq$ 8M$_{\odot}$) is 
involved if Be production is due to acceleration of the debris of grains 
formed in the ejecta of (in this case) SN~II. Determining the [Be/Fe] ratio 
in very metal-poor objects is then the key not only to fully test the 
hypothesis of a possible Big Bang Be production, but also to disentangle 
between the two other possibilities: in the first case [Be/Fe] is predicted 
to be enhanced at very early times, in the second case is constant. Of 
course, in order to do that, several new accurate measurements are needed.

\section{The Lithium Abundance of CD$-$24\degr17504: An Important 
Data Point}

As summarized in Table~1, CD$-$24\degr17504 was observed in different
instrumental settings, thus a larger spectral coverage is available, 
including the region around the \ion{Li}{i} line at 6708~\AA, which appears on 
both UVES Red settings (Red 580~nm and Red 860~nm). The coadded spectrum of 
the \ion{Li}{i} region has a total S/N ratio of $\sim 270$. Due to its low 
metallicity the star plays a crucial role in connection to the possible 
presence of a dependence of the Li abundance on metallicity in the Spite 
plateau.

In the combined spectrum, shown in Fig.~\ref{fig2}, the EW of the Li line is 
found to be EW~=~$20.57\pm 0.58$ m\AA, where the error bar was estimated with 
the Cayrel formula (Cayrel 1988). This value is consistent with the mean 
value obtained by measurements on the individual spectra ($20.44 \pm 0.56$). 
Previous measurements in the literature show a wide scatter between 
$15.1 \pm 2.3$ m\AA~(Ryan et al. \cite{ryan99}) and $25.1 \pm 4.1$ m\AA~(Spite 
\& Spite \cite{sspite2}). The average of all the available measurements is in 
good agreement with our measurement. However, if we compare our result to 
the recent study of lithium by Ryan et al. (\cite{ryan99}), who adopt the 
value of $18.1 \pm 1.3$, which is the mean of three measurements, we measure 
an EW which is larger by 2.5~m\AA, though consistent at 1.7 $\sigma$. We 
computed the Li abundances as described in Bonifacio \& Molaro (\cite{bm97}), 
from ATLAS9 no--overshooting, $\alpha$ enhanced models and obtained 
A(Li)~=~$2.13 \pm 0.08$ for \teff~=~6300~K, and A(Li)~=~$2.19 \pm 0.08$ for 
\teff~=~$6373 \pm 102$~K. By correcting the latter value (in order to be 
consistent with the temperature scale used by Bonifacio \& Molaro 
\cite{bm97}) for the NLTE effects, according to Carlsson et al. 
(\cite{carlsson94}) we obtain A(Li)~=~2.20, which is in agreement, within 
errors, with the plateau level of $2.238 \pm 0.012$ derived by Bonifacio 
\& Molaro (\cite{bm97}) using the same technique and IRFM temperatures. 
Thus our measure does not support the decrease of Li abundance 
for lower metallicities as claimed by Ryan et al. (\cite{ryan99}). Although 
a full discussion of the slope on the Spite plateau is beyond the scope of 
the present paper, we note the upwards revision of the EW which implies a Li 
abundance 0.06~dex higher than the value derived by Ryan et al. 
(\cite{ryan99}). This slight increase would then bring the data point up 
again, closer to the average Spite-plateau Li value, weakening their claim 
for the existence of a slope. By performing ordinary least squares and 
BCES\footnote{Bivariate Correlated Errors and Intrinsic scatter (Akritas \& 
Bershady \cite{akri96}).} fits to the Ryan et al. data, increasing their Li 
abundance for CD$-$24\degr17504 by 0.06 dex, we find that the slope decreases 
by $\sim$ 19\% . Therefore the main effect remains the true effective 
temperature of this star and in general which is the ``best'' temperature 
scale for metal--poor stars.
  
\section{Concluding Remarks}
We have presented the analysis of two new high resolution and high S/N 
near-UV spectra, obtained during the Commissioning of UVES, with the main 
purpose of measuring Be. We have detected Be in LP 815--43 ([Be/H]~=~$-$13.09, 
99.7\% confidence level), whereas an upper limit was found in the case of 
CD$-$24$\degr$17504 ([Be/H]~$\leq-$~13.39). We have also measured lithium in 
CD$-$24$\degr$17504, and found to be A(Li)~=~2.20, in good agreement with the 
Spite-plateau level. 

These new observations clearly show the potential of the new ESO VLT 
high resolution echelle spectrograph UVES, especially in the near-UV spectral 
range. Its (now measured) efficiency at 3130~\AA~(where the \ion{Be}{ii} 
doublet falls) is a factor of 3 to 4 higher than the combination of Keck~I 
and HIRES. New accurate measurements in a large sample of targets are 
foreseen in the near future and will have an important impact on our 
knowledge of Galactic cosmic-ray spallation. 

\begin{acknowledgements}
The authors would like to thank the whole UVES Team and the Project 
Scientist Sandro D'Odorico (ESO) for the successful Commissioning of the 
instrument that offered the astronomical community the great opportunity to 
work on high quality scientific data. Special thanks go to the members 
of the UVES Science Team, P. E. Nissen, B. Gustafsson, H. 
Hensberge, and P. Molaro, for making UVES performances optimized in the 
ultraviolet domain. Comments and suggestions from an anonymous referee 
notably improved the manuscript.  
\end{acknowledgements}


\end{document}